\documentclass{article}

\usepackage{graphicx} 
\usepackage{url} 

\newcommand{\iftwocol}[2]{#2}
\newcommand{\mitbf}[1]{\mathbf{#1}}

\newcommand{\m}{\mathbf{m}} 
\newcommand{\data}{\mathbf{d}}
\newcommand{\W}{\mathbf{W}}
\newcommand{\A}{\mathbf{A}}
\newcommand{\w}{\mathbf{w}}
\newcommand{\I}{\mathbf{I}}
\newcommand{\e}{\mathbf{e}}
\newcommand{\bx}{\mitbf{x}} 
\newcommand{\tr}{T} 
\newcommand{\dd}{\mathrm{d}} 
\newcommand{\degr}{\ensuremath{{}^\circ}}

\title{Tomographic inversion using $\mathbf{\ell_1}$-norm regularization of wavelet coefficients}
\author{Ignace Loris${}^{1,2}$, Guust Nolet${}^3$, Ingrid Daubechies${}^1$, F. A. Dahlen${}^3$\\
${}^1$Program in Applied and Computational Mathematics,\\ Princeton
University, Princeton, New Jersey, USA\\
${}^2$Dienst Theoretische Natuurkunde,
Vrije Universiteit Brussel,\\ Brussels, Belgium, E-mail: igloris@vub.ac.be\\
${}^3$Department of Geosciences, Princeton University, Princeton, New Jersey, USA}

\begin{document}

\maketitle

\begin{abstract}
We propose the use of $\ell_1$ regularization in a wavelet basis for
the solution of linearized seismic tomography problems $\A\m=\data$,
allowing for the possibility of sharp discontinuities superimposed on a smoothly varying background.
An iterative method is used to find a sparse solution $\m$ that contains no more
fine-scale structure than is necessary to fit the data $\data$ to within its assigned errors.

keywords: inverse problem, one-norm, sparsity, tomography, wavelets
\end{abstract}

\section{Introduction}

Like most geophysical inverse problems, the linearized problem
$\A\m=\data$ in seismic tomography is underdetermined, or at best
offers a mix of overdetermined and underdetermined parameters. It
has therefore long been recognized that it is important to suppress
artifacts that could be falsely interpreted as `structure' in the
earth's interior. Not surprisingly, strategies that yield the
smoothest solution $\m$ have been dominant in most global or
regional tomographic applications; these strategies include seeking
global models represented as a low-degree spherical harmonic
expansion \cite{dziewonski75,dziewonski87,masters96} as well as
regularization via minimization of the gradient ($\nabla \m$) or
second derivative ($\nabla^2 \m$) norm of a dense local
parametrization
\cite{nolet87,constable87,spakman88,vandecar94,trampert96}.

Smooth solutions, however, while not introducing small-scale
artifacts, produce a distorted image of the earth through the strong
averaging over large areas, thereby making small-scale detail
difficult to see, or even hiding it. Sharp discontinuities are
blurred into gradual transitions. For example, the inability of
global, spherical-harmonic, tomographic models to yield as clear an
image of upper-mantle subduction zones as produced by more localized
studies has long been held against them. \cite{deal99} and
\cite{deal99a} optimize images of upper-mantle slabs to fit physical
models of heat diffusion, in an effort to suppress small-scale
imaging artifacts while retaining sharp boundaries.
 \cite{portniaguine99} use a
conjugate-gradient method to seek the smallest possible anomalous
domain by minimizing a norm based on a renormalized gradient $\nabla
\m / ( \nabla \m \cdot \nabla \m + \gamma^2 )^\frac12$, where
$\gamma$ is a small constant. Like all methods that deviate from a
least-squares type of solution, both these methods are nonlinear and
pose their own problems of practical implementation.

The notion that we should seek the `simplest' model $\m$ that fits a
measured set of data $\data$ to within the assigned errors is
intuitively equivalent to the notion that the model should be
describable with a small number of parameters. But, clearly,
restricting the model to a few low-degree spherical-harmonic or
Fourier coefficients, or a few large-scale blocks or tetrahedra,
does not necessarily lead to a geophysically plausible solution. In
this paper we investigate whether a multiscale representation based
upon wavelets \cite{daubechies92} has enough flexibility to
represent the class of models we seek. We propose an $\ell_1$-norm
regularization method which yields a model $\m$ that has a strong
tendency to be {\it sparse} in a wavelet basis, meaning that it can
be faithfully represented by a relatively small number of nonzero
wavelet coefficients. This allows for models that vary smoothly in
regions of limited coverage without sacrificing any sharp or
small-scale features in well-covered regions that are required to
fit the data. Our approach is different from an approach briefly
suggested by \cite{deHoop2005}, in which the mapping between data
and model is decomposed in curvelets: here we are concerned with
applying the principle of parsimony to the solution of the inverse
problem,
 without any special preference for singling out linear
features, for which curvelets are probably better adapted than wavelets.

In Section 2 we give a short description of the mathematical method,
and in Section 3 we consider a geophysically motivated, toy 2D
application, in which the synthetic data are a small set of
regional, fundamental-mode, Rayleigh-wave dispersion measurements
expressed as wavenumber perturbations $\delta k(\nu )$ at various
frequencies $\nu$. To enable us to concentrate on the mathematical
rather than the geophysical aspects of the inverse problem, we
assume that the fractional shear-velocity perturbations
$\delta\hspace{-0.1em}\ln\hspace{-0.1em}\beta=\delta\beta/\beta$
within the region are depth-independent. Finite-frequency
interpretation of the surface-wave dispersion data  \cite{zhou04}
then yields a 2D linearized inverse problem of the form $\A  \m =
\data$. We compare wavelet-basis models $\m$ obtained using our
proposed $\ell_1$-norm regularization with models obtained using
more conventional $\ell_2$ regularization, both with and without
wavelets, and show that the former are sparser and have fewer
small-scale artifacts.

\section{Mathematical principles}

In any realistic tomographic problem, the linear system $\A\m=\data$
is not invertible: even when the number of data exceeds the number
of unknowns, the least-squares matrix $\A^\tr\A$ is (numerically)
singular. Additional conditions always have to be imposed. The
proposed regularization method is based on the fundamental
assumption that the model $\m$ is sparse in a wavelet basis
\cite{daubechies92}. We believe that this is an appropriate
inversion philosophy for finding a smoothly varying model while
still allowing for whatever sharp or small-scale features are
required to fit the data $\data$. An important feature of the method
is that the location of the small-scale features does not have to be
specified beforehand.

A wavelet decomposition is a special kind of basis transformation
that can be computed efficiently (the number of operations is
proportional to the number of components in the input). At each step
the algorithm strips off detail belonging to the finest scale
present ---this detail is encoded in wavelet coefficients, broadly
corresponding to local differences--- and calculates a coarse
version ---encoded in scaling coefficients, broadly corresponding to
local averages--- that is only half the size of the original in 1D
and only one quarter the size in 2D. This procedure is repeated on
the successive coarse versions.  The resulting wavelet coefficients
(at the different scales) and scaling coefficients (at the final
coarsest scale only) are called the wavelet decomposition of the
input.  By this construction each wavelet coefficient carries
information belonging to a certain scale (by virtue of the
decimation) and a certain position (use of \emph{local}
differences). The final few scaling coefficients represent a (very)
coarse average.

The mathematical relation between the wavelet-basis
expansion coefficients $\w $ and the model $\m$ is
the wavelet transform $\W$ (a linear operator): $\m=\W^\tr \w $.
By choosing the local differences and averages
carefully (corresponding to a choice among many different so-called
wavelet families), the inverse transformation from $\w$ back to $\m$
can be made equally efficient.
In our application we will use a special kind of 2D wavelet basis that is
overcomplete: it contains six different wavelets corresponding to
different directions. Because of this overcompleteness,
the wavelet transform $\W$ has a left inverse
(namely $\W^\tr$): $\W^\tr \W=\I$, but no right inverse, $\W\W^\tr\neq \I$.
Appendix~\ref{waveletappendix} contains a short overview of this particular
construction. In short our wavelet and scaling coefficients $\w$ contain
information on scale, position and direction.

\begin{figure*}
\centering\includegraphics{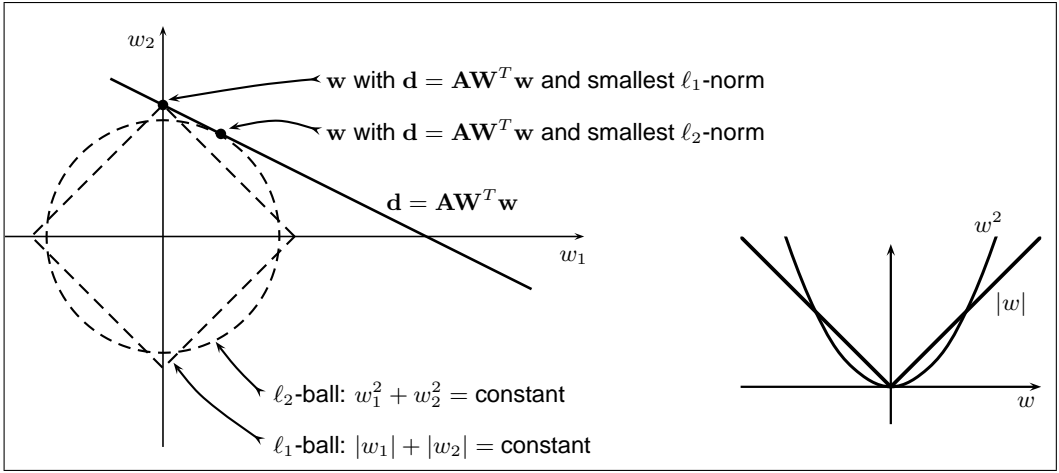}

\caption{Sparsity, $\ell_1$ minimization and $\ell_2$ minimization:
Left: Because the $\ell_1$-ball has no bulge, the solution with
smallest $\ell_1$-norm is sparser than the solution with smallest
$\ell_2$-norm. Right: A $|w|$ penalization effects small
coefficients more and large coefficients less than the (traditional)
$w^2$ penalization.}\label{diamondpic}
\end{figure*}

For the tomographic reconstruction, we will require a sparse set of
wavelet-basis coefficients: the vast majority of these represent
differences and will only be present around non-smooth features. In
this way we regularize the inversion by adapting ourselves to the
model rather than to the operator. As a measure of sparsity we will
use the $\ell_1$-norm of the wavelet representation $\w $ of the
model $\m$, i.e. we will look for a solution of the linear equations
$\A\m=\data$ that has a small $\|\w \|_1=\sum_i |w_i|$. Since
$|w_i|>|w_i|^2$ for small $w_i$ and $|w_i|<|w_i|^2$ for large $w_i$,
this type of penalization will favor a small number of large
coefficients over a large number of small coefficients in the
reconstruction (whereas a traditional $\ell_2$ penalization might do
the opposite).  We are not claiming that the sparsest solution
always coincides with the minimum $\ell_1$-norm solution, but one
can show that it often does \cite{Donoho2006,Candes}. A schematic
justification for this is given in Fig.~\ref{diamondpic}.

In particular, our strategy will consist of searching for the
minimizer of the functional
\begin{equation}
I_1(\w)=\|\data-\A \m\|_2^2+2\tau \|\w \|_1
=\|\data-\A \W^\tr\w\|_2^2+2\tau \|\w \|_1 ,
\label{l1functional}
\end{equation}
where $\tau$ is an adjustable parameter at our disposal.
Here, the first (quadratic) term corresponds to the conventional
statistical measure of misfit to the data,
$\chi^2=\sum_i[d_i-(\A\m)_i]^2$, and the second
($\ell_1$-norm) term $2\tau\|\w\|_1$ is introduced to regularize the inversion.
In writing $\chi^2$ in this form, we have made the simplifying
assumption that the noisy data $\data$ are uncorrelated with unit variance.
More generally, the misfit portion of the functional~(\ref{l1functional})
is $\chi^2=(\data-\A\m)^\tr\mathbf{\Sigma}^{-1}(\data-\A\m)$, where
$\mathbf{\Sigma}$ is the data covariance matrix. In the 2D toy problem
considered in Section~3, we invert synthetic data $\data$ having a
constant (but non-unit) variance, $\mathbf{\Sigma}=\sigma^2\I$.

The minimizer of the functional (\ref{l1functional}) can be found by
iteration \cite{DDD}: starting with the present approximation
$\w^{(n)}$ one constructs an $n$th-iterate surrogate functional
\begin{equation}
I_1^{(n)}(\w)=I_1(\w)-\|\A
\W^\tr(\w-\w^{(n)})\|_2^2+\|\w-\w^{(n)}\|_2^2
\end{equation}
that has the same value and the same derivative at the point
$\w=\w^{(n)}$ as the original functional (see Fig.
\ref{iterationpic}). This surrogate functional can be rewritten as
\iftwocol{\begin{equation}\begin{array}{l} \displaystyle
\!\!I_1^{(n)}(w)=\left\|\w\!-\!\left(\W\A^\tr
\data+(\I-\W\A^\tr \A \W^\tr)\w^{(n)}\right)\right\|_2^2\\[2mm]
\displaystyle
\qquad\qquad\qquad\qquad\qquad\qquad\qquad\qquad\,\,+2\tau
\|\w\|_1+c^{(n)},\end{array} \label{newfunctional}
\end{equation}}
{\begin{equation} I_1^{(n)}(w)=\left\|\w-\left(\W\A^\tr
\data+(\I-\W\A^\tr \A \W^\tr)\w^{(n)}\right)\right\|_2^2+2\tau
\|\w\|_1+c^{(n)}, \label{newfunctional}
\end{equation}} where $c^{(n)}$ is independent of $\w$.  This
functional has a much simpler form than the original $I_1(\w)$
because there is no operator $\A\W^\tr$ mixing different components
of $\w$. The next approximation $\w^{(n+1)}$ is defined by the
minimizer of this new functional. By calculating the derivative of
expression (\ref{newfunctional}) with respect to a specific wavelet
or scaling coefficient $w_i$, one finds the following set of
component-by-component equations:
\begin{equation}
w_i-\left(\W\A^\tr \data+(\I-\W\A^\tr \A
\W^\tr)\w^{(n)}\right)_i+\tau\, \mathrm{sign}(w_i)=0,
\end{equation}
valid whenever $w_i\neq 0$. These equations are solved by
distinguishing the two cases $w_i>0$ and $w_i<0$;  the solution
---corresponding to the minimizer of the surrogate functional
$I_1^{(n)}(w)$, and denoted by $\w^{(n+1)}$--- is then found to
equal
\begin{equation}
\w ^{(n+1)}=\mathcal{S}_\tau\!\left[\W \A ^\tr \data+(\I -\W \A ^\tr
\A \W^\tr)\w ^{(n)}\right],
\label{iterationformula}
\end{equation}
where $S_\tau$ is the so-called soft-thresholding operation, i.e.
\begin{equation}
\mathcal{S}_{\tau}(w)=\left\{
\begin{array}{lllcl}
w-\tau & w & \geq & \tau\\
0 & |w| &\leq &\tau\\
w+\tau\qquad\qquad & w & \leq& -\tau,
\end{array}\right.
\end{equation}
performed on each wavelet or scaling coefficient $w_i$ individually.
The starting point of the iteration procedure is arbitrary, e.g. $\w
^{(0)}=\mathbf{0}$. Because of the component-wise character of the
tresholding, it is straightforward to use different thresholds
$\tau_i$ for different components $w_i$ if desired, and in fact we
shall use different thresholds $\tau_{\rm w}$ and $\tau_{\rm s}$ for
the wavelet and scaling coefficients in our application. A schematic
representation of the idea behind the iteration
(\ref{iterationformula}) is given in Fig.\ \ref{iterationpic}. We
realize that this iteration converges slowly for ill-conditioned
matrices, but we use it here because it is proven to converge to the
solution \cite{DDD}.

\begin{figure*}
\centering\resizebox{\textwidth}{!}{\includegraphics{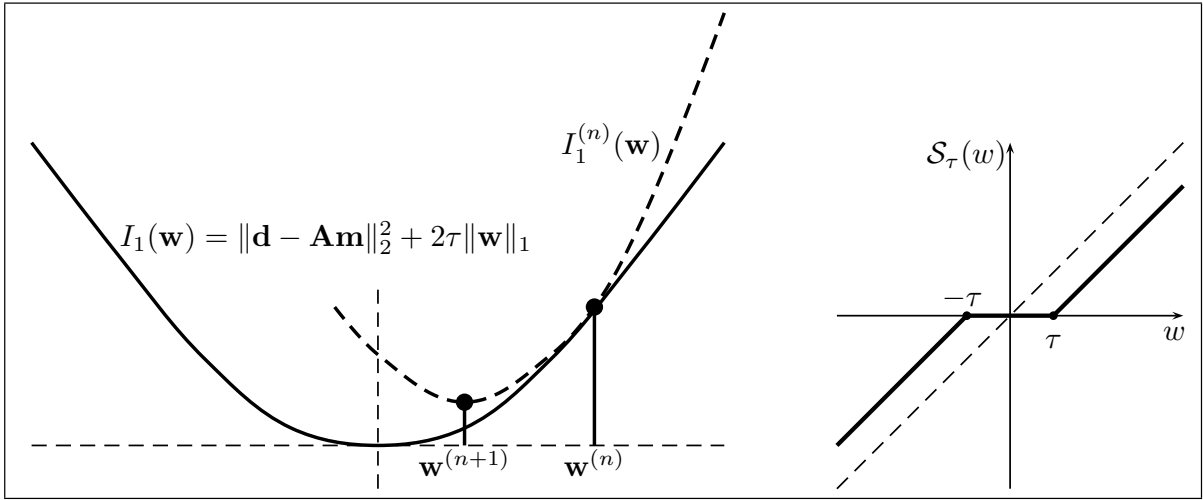}}

\caption{Left: The functional $I_1(\w)$ is approximated in the vicinity of
$\w^{(n)}$ by a surrogate functional $I_1^{(n)}(\w)$,
constructed in such a way that its minimum is easy to find (eq.\
(\ref{iterationformula})). This defines the next step in the
iteration. Right: Soft thresholding function $\mathcal{S}_{\tau}(w)$.}
\label{iterationpic}
\end{figure*}

An improvement in convergence can be gained by rescaling the
operator $\A $ (and rescaling the data $\data$ at the same time) in
such a way that the largest eigenvalue of $\alpha^2 \A ^\tr \A $ is
close to (but smaller than) unity. The iteration corresponding to the
minimization of this new, rescaled functional is
\begin{equation}
\w ^{(n+1)}=\mathcal{S}_{\tau\alpha^2}\!\left[\alpha^2\W \A ^\tr
\data+(\I -\alpha^2\W \A ^\tr \A  \W^\tr)\w ^{(n)}\right].
\label{improvediteration}
\end{equation}
We will also make use of the  following two-step procedure: from the
outcome $\overline{\m}=\W^\tr \overline{\w }$ of the iteration
(\ref{improvediteration}), we define new, linearly shifted data
$\data'=2\data-\A \overline{\m}$ and restart the same iteration with
this new data: \iftwocol{\begin{equation} \label{tonyiter1} \w
^{(n+1)}=\mathcal{S}_{\tau\alpha^2}\!\left[\alpha^2\W \A ^\tr
\data'+(\I -\alpha^2\W \A ^\tr \A \W^\tr)\w ^{(n)}\right],
\end{equation}$\w ^{(0)}=\overline{\w}$.}
{\begin{equation} \label{tonyiter1} \w
^{(n+1)}=\mathcal{S}_{\tau\alpha^2}\!\left[\alpha^2\W \A ^\tr
\data'+(\I -\alpha^2\W \A ^\tr \A \W^\tr)\w
^{(n)}\right],\qquad\qquad \w ^{(0)}=\overline{\w }.
\end{equation}} The outcome
$\overline{\overline{\m}}=\W^\tr \overline{\overline{\w }}$ of this
second iteration is then the final, regularized reconstruction of
the model. For the same value of the regularization parameter
$\tau$, the second step improves the data fit considerably,
$\|\data-\A \overline{\overline{\m}}\|_2^2<\|\data-\A
\overline{\m}\|_2^2$; hence a given level of final data fit $\chi^2$
will, in the two-step procedure, correspond to a higher value of
$\tau$. Because $\tau$ specifies the threshold level, a higher value
will lead to more aggressive thresholding and thus faster
convergence to a sparse solution.

The above method will be demonstrated in the next section and
compared to a conventional $\ell_2$-regularization method, in which
the functional
\begin{equation}
I_2(\m)=\|\data-\A \m\|_2^2+\tau \|\m\|_2^2
\end{equation}
is minimized (the crucial difference with $I_1(\w )$ being the
second term). This gives rise to the familiar system of damped
normal equations
\begin{equation}
(\A ^\tr \A +\tau \I )\m=\A^\tr\data, \label{l2regsys}
\end{equation}
whose solution $\m=(\A ^\tr \A +\tau \I )^{-1}\A^\tr\data$ can be
found using a linear solver of choice, since $\A^\tr \A +\tau\I$ is
a regular matrix. To emphasize the similarities and differences with
the $\ell_1$ method, we adopt the classical Landweber iteration
\cite{landweber} that can be (but in modern applications seldom is)
used for solving the linear equations (\ref{l2regsys}):
\iftwocol{\begin{equation} \m^{(n+1)}=\A ^\tr \data+\left[\I -(\A
^\tr \A +\tau \I )\right]\!\m^{(n)},\qquad \m^{(0)}=\mathbf{0}.
\end{equation}}{\begin{equation} \m^{(n+1)}=\A ^\tr \data+\left[\I -(\A
^\tr \A +\tau \I )\right]\!\m^{(n)},\qquad\qquad
\m^{(0)}=\mathbf{0}.
\end{equation}}
No thresholding is employed here. Rescaling of the operator and the
data again improves the rate of convergence:
\iftwocol{\begin{equation} \m^{(n+1)}=\alpha^2\A ^\tr\data+\left[\I
-(\alpha^2\A ^\tr \A +\tau\alpha^2 \I ) \right]\!\m^{(n)},
\label{linearLW}
\end{equation} $\m^{(0)}=\mathbf{0}$.}{\begin{equation} \m^{(n+1)}=\alpha^2\A
^\tr\data+\left[\I -(\alpha^2\A ^\tr \A +\tau\alpha^2 \I )
\right]\!\m^{(n)},\qquad\qquad \m^{(0)}=\mathbf{0}. \label{linearLW}
\end{equation}}
Of course it is also possible to solve the linear
system~(\ref{l2regsys}) using a conjugate-gradient or similar
algorithm in much less time.

A third option is to use an $\ell_2$ penalization on the wavelet
coefficients. This allows us to penalize the scaling coefficients
differently than the wavelet coefficients (with the help of
different penalization parameters $\tau_{\rm s}$ and $\tau_{\rm
w}$). We can use the following iteration, similar to formula
(\ref{linearLW}), but now in the wavelet domain:
\iftwocol{\begin{equation} \label{tonyiter2} \w^{(n+1)}=\alpha^2\W\A
^\tr\data+\left[\I -(\alpha^2\W\A ^\tr \A\W^\tr +\alpha^2 \tilde\I )
\right]\!\w^{(n)},
\end{equation}$\w^{(0)}=\mathbf{0}$,}{\begin{equation} \label{tonyiter2}
\w^{(n+1)}=\alpha^2\W\A ^\tr\data+\left[\I -(\alpha^2\W\A ^\tr
\A\W^\tr +\alpha^2 \tilde\I ) \right]\!\w^{(n)},\qquad\qquad
\w^{(0)}=\mathbf{0},
\end{equation}}
where $\tilde\I$ acts as the $\tau_{\rm w}\times\mbox{identity}$ on
wavelet coefficients and as the $\tau_{\rm s}\times\mbox{identity}$
on scaling coefficients.
 If we were to use an orthonormal wavelet basis ($\W^\tr\W=\W\W^\tr=\I$) for our
expansions, and if we penalized every coefficient the same,
$\tau_{\rm w}=\tau_{\rm s}=\tau$, then this method would be identical
to the previous $\ell_2$ method.

In the following we consider both one-step and two-step $\ell_1$
wavelet penalization as well as conventional $\ell_2$ penalization,
both without and with wavelets, using the rescaled iterative
schemes~(\ref{improvediteration}), (\ref{tonyiter1}),
(\ref{linearLW}) and~(\ref{tonyiter2}) for the purposes of
comparison.

\section{Implementation}
\label{implementationsection}

To test the above ideas, we devised a dramatically simplified,
two-dimensional, synthetic surface-wave inversion problem very
loosely modeled after an actual \textsc{Passcal} deployment in
Tanzania \cite{owens95}. Fig.~\ref{modelfigure} (left) shows the
hypothetical experimental setup: the highly schematized input model
consists of a sharp, bent, East African rift structure with low
shear-wave velocity,
$\delta\hspace{-0.1em}\ln\hspace{-0.1em}\beta(x,y)<0$, superimposed
upon a smooth, circular cratonic positive anomaly,
$\delta\hspace{-0.1em}\ln\hspace{-0.1em}\beta(x,y)>0$. Eleven
earthquake events (circles) were taken from the NEIC catalogue to
mimic realistic regional seismicity for the duration of a typical
temporary deployment of the twenty-one stations (triangles). The
locations of the seismic stations and events are listed in
Table~\ref{stationtable}. For each of the $11\times 21$
source-receiver paths, we assume that fundamental-mode Rayleigh-wave
perturbations $\delta k(\nu)$ have been measured at eight selected
frequencies between $\nu\approx 0.01$~Hz and $\nu\approx 0.1$~Hz.
These wavenumber perturbations are related to the 2D,
depth-independent velocity perturbations
$\delta\hspace{-0.1em}\ln\hspace{-0.1em}\beta(x,y)$ via a 2D,
frequency-dependent sensitivity kernel (see Appendix A for more
details):
\begin{equation}
\delta k (\nu ) = \int\!\!\!\int K_{\rm 2D}(x,y,\nu )\,
\delta\hspace{-0.1em}\ln\hspace{-0.1em}\beta(x,y ) \, \dd x\, \dd y \label{tonykernel}.
\end{equation}
Plots of the lowest-frequency ($\nu\approx 0.01$~Hz) and
highest-frequency ($\nu\approx 0.1$~Hz) kernel $K_{\rm 2D}(x,y,\nu)$
for a typical source-receiver pair are shown in the left two panels
of Fig.~\ref{coveragepic}. Because finite-frequency scattering and
diffraction effects are accounted for in the kernels $K_{\rm
2D}(x,y,\nu)$, there is significant off-path sensitivity of the
measurements $\delta k(\nu)$ within the first one or two Fresnel
zones \cite{zhou04}. All kernels $K_{\rm 2D}(x,y,\nu)$ and distances
are computed in the flat-earth earth approximation.

The study region, which is $35^{\circ}$ (north-south) by
$25^{\circ}$ (east-west), is subdivided into $N_x\times N_y
=64\times 64=4096$ equal-sized rectangles, and the discretized model
vector $\m$ consists of the unknown constant values of
$\delta\hspace{-0.1em}\ln\hspace{-0.1em}\beta(x,y )$ within each
rectangle. To compute the matrix $\A$, which maps the discretized
model $\m$ onto the data $\data$ (consisting of multiple $\delta
k(\nu)$), each kernel $K_{\rm 2D}(x,y,\nu)$ is sampled $n_x\times
n_y$ times on each of the $N_x\times N_y$ model-vector rectangles
and a Riemann sum is used to compute the quantity
\iftwocol{\begin{equation}\label{tonyapproxint}
\begin{array}{l}
\displaystyle \int\!\!\!\int_{\mathrm{rectangle}
(k,l)}\hspace{-2.0em} K_{\rm 2D}(x,y,\nu)\,\dd x\,\dd y\approx\frac{\Delta x\Delta
y}{n_xn_y}\\[3mm]
\displaystyle \qquad\qquad\times\sum_{m,n}K_{\rm
2D}\left(x_l-\frac{\Delta x}{2} +(m-\frac{1}{2})\delta x,\right.\\
\displaystyle \qquad\qquad\qquad\qquad\qquad\quad\left.
y_k-\frac{\Delta y}{2}+(n-\frac{1}{2})\delta y,\,\nu\right),
\end{array}
\end{equation}}{\begin{equation}
\label{tonyapproxint} \int\!\!\!\int_{\mathrm{rectangle}
(k,l)}\hspace{-2.0em} K_{\rm 2D}(x,y,\nu)\,\dd x\,\dd y\approx
\frac{\Delta x\Delta y}{n_xn_y}\sum_{m,n}K_{\rm 2D}\left(x_l-\Delta
x/2 +(m-1/2)\delta x,\,y_k-\Delta y/2+(n-1/2)\delta y,\,\nu\right),
\end{equation}}
where $\delta x=\Delta x/n_x$ and $\delta y=\Delta y/n_y$. A
schematic representation of the $N_x\times N_y$ grid on which the
spatial-domain model $\m$ is specified and the $n_x\times n_y$
integration subgrid is shown in Fig.~\ref{gridfigure}. We choose
$n_x=n_y=32$ since we have found that doubling this to $n_x=n_y=64$
yields a change of less than one percent in the integrated value of
$\A$. The dimensions of the resulting matrix $\A$ are 1848 (number
of stations $\times$ number of events $\times$ number of
wavenumbers) by 4096 (number of model-vector pixels). To give an
idea of the overall degree of coverage, we have plotted the sum
(over all station event pairs) of the absolute value of all of the
lowest-frequency and all the highest-frequency discretized kernels
in the right two panels of Fig.~\ref{coveragepic}. It is clear that
much of the study area, particularly in the northwest and southeast,
is completely uncovered (as is typical of real-world, regional
seismic experiments).

\begin{figure*}
\centering\resizebox{\textwidth}{!}{\includegraphics{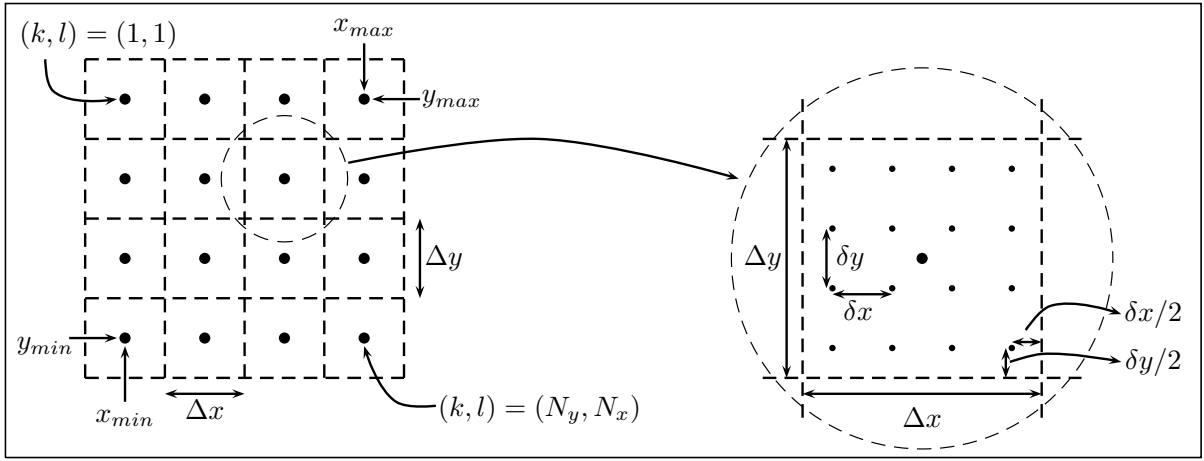}}

\caption{Schematic representation of the 2D Cartesian grids used.
Left: $N_x\times N_y$ grid used to specify the model $\m$. Right:
Blowup of the finer-scale $n_x\times n_y$ grid used to compute the
kernel matrix $\A$ via the approximate
integration~(\ref{tonyapproxint}). Since the study region is
rectangular in shape (see Fig.~\ref{modelfigure}) and since $N_x=
N_y$ and $n_x= n_y$, the actual $\Delta x\times\Delta y$ model
pixels and $\delta x\times \delta y$ integration subpixels are also
rectangular, rather than square as shown.} \label{gridfigure}
\end{figure*}

\begin{table}
\caption{List of positions of seismic stations and earthquake events used in the
synthetic inversion.}
 \centering
\begin{tabular}{|r@{.}lr@{.}lcr@{.}lr@{.}l|}
\hline
\multicolumn{4}{|c||}{Stations}&&\multicolumn{4}{|c|}{Events}\\
 \hline \multicolumn{2}{c}{longitude} & \multicolumn{2}{c}{latitude}
 &&
 \multicolumn{2}{c}{longitude} & \multicolumn{2}{c}{latitude}\\ \hline
$33$&$3203\degr$ &  $-7$&$9073\degr$ && $29$&$02\degr$ & $-1$&$86\degr$ \\
$35$&$1382\degr$ &  $-4$&$3238\degr$ && $49$&$10\degr$ & $12$&$85\degr$\\
$32$&$7712\degr$ &  $-9$&$2958\degr$ && $44$&$15\degr$ & $11$&$80\degr$\\
$33$&$2588\degr$ &  $-8$&$1060\degr$ && $30$&$82\degr$ & $-7$&$84\degr$\\
$29$&$6927\degr$ &  $-4$&$8392\degr$ && $46$&$34\degr$ & $12$&$33\degr$\\
$38$&$6170\degr$ &  $-5$&$3018\degr$ && $39$&$17\degr$ & $19$&$02\degr$\\
$30$&$3988\degr$ &  $-5$&$1168\degr$ && $32$&$78\degr$ & $5$&$06\degr$\\
$36$&$5695\degr$ &  $-5$&$3223\degr$ && $44$&$15\degr$ & $14$&$57\degr$\\
$37$&$4763\degr$ &  $-5$&$3775\degr$ && $28$&$84\degr$ & $1$&$16\degr$\\
$36$&$7192\degr$ &  $-3$&$8422\degr$ && $40$&$33\degr$ & $14$&$20\degr$\\
$35$&$7965\degr$ &  $-4$&$9040\degr$ && $33$&$67\degr$ & $-3$&$05\degr$\\
$36$&$6983\degr$ &  $-2$&$7252\degr$ \\
$34$&$3462\degr$ &  $-4$&$9610\degr$ \\
$34$&$0560\degr$ &  $-6$&$0192\degr$ \\
$35$&$4007\degr$ &  $-5$&$2508\degr$ \\
$33$&$2415\degr$ &  $-8$&$9835\degr$ \\
$33$&$1842\degr$ &  $-4$&$7145\degr$ \\
$33$&$5180\degr$ &  $-6$&$9372\degr$ \\
$34$&$7315\degr$ &  $-4$&$6403\degr$ \\
$36$&$0163\degr$ &  $-3$&$8892\degr$ \\
$32$&$0832\degr$ &  $-5$&$0878\degr$ \\ \hline
\end{tabular}
\label{stationtable}
\end{table}

Using the matrix $\A$ and the input model $\m^{\mathrm{input}}$ with
a sharp, low-velocity East African rift superimposed on a broad,
high-velocity cratonic structure, we compute synthetic data
$\data=\A\m^{\mathrm{input}}+\e$, where we have added Gaussian noise
$\e$ with zero mean and a standard deviation equal to two percent of
the largest synthetic wavenumber perturbation, i.e.
$\sigma=0.02\,\max(|\A\m^{\mathrm{input}}|)$. By adopting a constant
standard deviation $\sigma$, errors at the highest frequency $\nu$
and unperturbed wavenumber $k(\nu)$ are more than an order of
magnitude smaller than those for the lowest frequency and wavenumber
data, where the signal-to-noise ratio may be close to unity. Since
finite-frequency inversions include the effect of scattered wave
energy, a high precision of the measurement $\delta k(\nu)$ at high
frequency $\nu$ is realistic. The purpose of the proposed algorithm
is now to reconstruct $\m$ from the knowledge of the noisy data
$\data$, the matrix $\A $ and the linear equations $\A\m=\data$.

\begin{figure*}
\centering\resizebox{\textwidth}{!}{\includegraphics{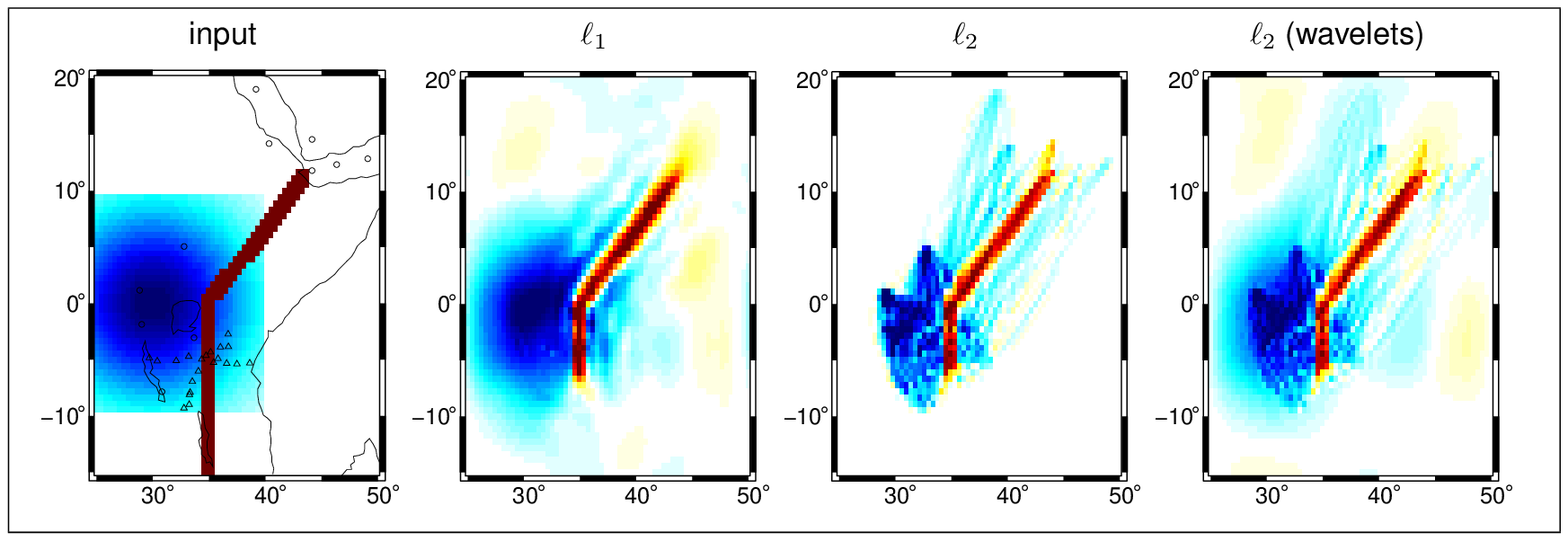}}

\caption{From left to right: Toy 2D velocity model for the East
African rift and adjacent continental craton, showing the seismic
stations (triangles) and earthquake events (circles); reconstructed
model using the two-step $\ell_1$-penalization method;
reconstruction using the spatial-domain $\ell_2$ method;
reconstruction using the wavelet-domain $\ell_2$ method. The
two-step $\ell_1$ model is that obtained after $1000+1000$
iterations, whereas both $\ell_2$ models are after 2000 Landweber
iterations. Red denotes low anomalous velocity,
$\delta\hspace{-0.1em}\ln\hspace{-0.1em}\beta(x,y)<0$, and blue
denotes high velocity,
$\delta\hspace{-0.1em}\ln\hspace{-0.1em}\beta(x,y)>0$. The absolute
magnitude $|\delta\hspace{-0.1em}\ln\hspace{-0.1em}\beta(x,y)|$ is
irrelevant, since the inverse problem $\A\m=\data$ is linear and the
synthetic data are constructed from the input model
$\m^{\mathrm{input}}$ (leftmost map) via
$\data=\A\m^{\mathrm{input}}+\e$.} \label{modelfigure}
\end{figure*}

\begin{figure*}
\centering\resizebox{\textwidth}{!}{\includegraphics{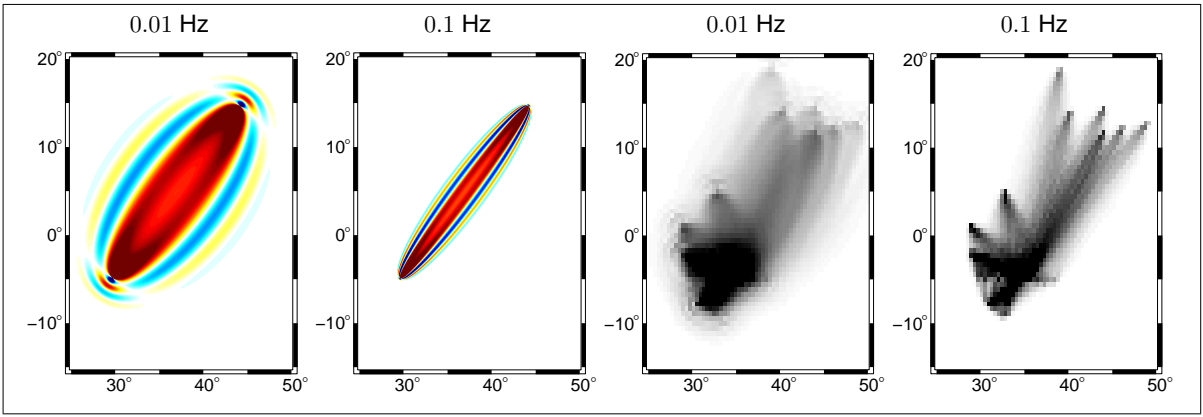}}

\caption{Left: Map view of a typical two-dimensional sensitivity
kernel $K_{\rm 2D}(x,y,\nu)$ at the lowest frequency considered,
$\nu\approx 0.01$~Hz. Second from left: Highest-frequency
($\nu\approx 0.1$~Hz) kernel for the same source-receiver path. Both
kernels exhibit structure on a much finer scale than the resolution
of the model, necessitating the $n_x\times n_y$ numerical
integration to compute the matrix $\A$ in eq.~(\ref{tonyapproxint}).
Red denotes negative values, $K_{\rm 2D}(x,y,\nu)<0$, and blue
denotes positive values, $K_{\rm 2D}(x,y,\nu)>0$. The cross-path
tapering of the kernels as a result of the finite time-domain taper,
eqs~(\ref{tony2})--(\ref{tony3}), is clearly visible. Second from
right: The sum (over all source-receiver pairs) of the absolute
value of the lowest frequency ($\nu\approx 0.01$~Hz) integrated
kernels (as computed in eq.~\ref{tonyapproxint}). Far right: The sum
(over all source-receiver pairs) of the absolute value of the
highest frequency ($\nu\approx 0.1$~Hz) integrated kernels (as
computed in eq.~\ref{tonyapproxint}). The coverage is adequate in
the vicinity of the East African rift (by design of the original
seismic deployment) but poor elsewhere.} \label{coveragepic}
\end{figure*}

\begin{figure*}
\centering\resizebox{\textwidth}{!}{\includegraphics{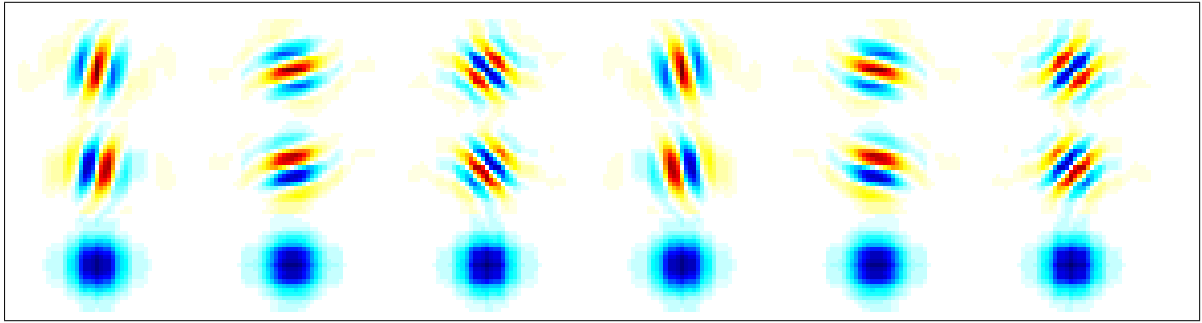}}

\caption{Spatial-domain structure of the 2D dual-tree complex
wavelets used in the reconstruction (figure taken/adapted from
\cite{selesnick06}). First row: real part, second row: imaginary
part, third row: norm squared (i.e. sum of the squares of the top
two plots). The directional character of each of the six wavelet
functions is clear. Four different wavelet scales, i.e. four
different replicas of this picture, each a factor of two smaller
than the one above it, are used in both the $\ell_1$ and $\ell_2$
wavelet-basis inversions.} \label{waveletpicb}
\end{figure*}

For our purposes we will make use of the overcomplete 2D wavelet
basis described by \cite{Kingsbury} and \cite{selesnick06} because
of its ability to distinguish different directions (see Fig.
\ref{waveletpicb}). We use four wavelet scales, for a total of
$4\times 64^2=16\,384$ wavelet and scaling coefficients $\w$ (four
times the number of model coefficients $\m$). The starting point for
the iterations in both the $\ell_1$ and $\ell_2$ inversions is $\w
=\mathbf{0}$ and $\m=\mathbf{0}$. As explained in the previous
section we renormalize the $\ell_1$ thresholded iteration by
choosing $\alpha=\lambda_{\mathrm{max}}^{-1/2}$ (which in our case
equals $4884.5$) where $\lambda_\mathrm{max}$ is the largest
eigenvalue of $\A ^\tr \A$. We let the iteration run for 1000 steps,
adjust the data (two-step procedure) and let the second-step
algorithm run for another 1000 steps. The threshold $\tau$ is chosen
by hand in such a way as to arrive at a final value for the
variance-adjusted misfit, $\chi^2=\|\data-\A \m\|_2^2\hspace{0.1
em}/\sigma^2$, that is approximately equal to 1848 (the number of
data). The noisy data $\data$ are thus fit to within their standard
errors $\sigma$ and no better; pushing the fit beyond this would
amount to fitting the noise $\e$, which would lead to undesirable
artifacts in the resulting model $\m$.

It should also be noted that the threshold\-ing is done on pairs of
wavelet coefficients: The wavelets come in pairs (at the same scale,
position and orientation) that we interpret as real and imaginary
part of a complex wavelet, i.e. thresholding corresponds to
$(w^\mathrm{re}_k,w^\mathrm{im}_k)\rightarrow z=w^\mathrm{re}_k+ i
w^\mathrm{im}_k\rightarrow \tilde z=z
\mathrm{S}_\tau(|z|)/|z|\rightarrow (\mathrm{Re}(\tilde
z),\mathrm{Im}(\tilde z))$. This particular method of thresholding
is borrowed from image denoising where it is found to make a big
difference in avoiding artifacts
\cite{guleryuz06,baraniuk03,selesnick05}. Furthermore, the threshold
for the diagonally oriented wavelets is multiplied by $1.2395$
because $\|\nabla\psi_{\pm 45^\circ}\|_1=
1.2395\|\nabla\psi_{\mathrm{other}}\|_1$. We choose the threshold
$\tau_{\mathrm{s}}$ for the scaling coefficients to be 1/10th of the
threshold $\tau_{\mathrm{w}}$ for the wavelet coefficients; since
the scaling coefficients correspond to a few large-scale averages
(64 in our case versus more than $16\,000$ finer-scale wavelet
coefficients) it is not so important that these be sparse. Likewise,
in the wavelet-basis $\ell_2$ inversions, we set the penalization
parameter for the scaling coefficients to 1/10th the value of the
penalization parameter for the wavelet coefficients.

The two-step $\ell_1$ algorithm takes about ten minutes for
$1000+1000$ iterations on a 1.5GHz PC. The result of the $\ell_1$
inversion is compared with the outcome of both of the $\ell_2$
methods, with and without using wavelets, with the thresholding or
penalization parameter $\tau$ chosen in every case to achieve the
same data fit: $\chi^2\approx 1848$ (see Fig. \ref{chi2pic}). The
number of Landweber iterations is 2000, so that the total number of
two-step $\ell_1$ and single-step $\ell_2$ iterations is the same.
The spatial-domain $\ell_2$-regularization method yields a relative
modeling error
$\|\m-\m^{\mathrm{input}}\|_2/\|\m^{\mathrm{input}}\|_2$ of about
$74\%$, whereas the two-step $\ell_1$ method yields a relative
modeling error of only $47\%$ (see Fig. \ref{errorpic}), and is
clearly less noisy (compare the middle two maps in
Fig.~\ref{modelfigure}). The wavelet-basis $\ell_2$-regularized
inversion (rightmost map in Fig.~\ref{modelfigure}) is only slightly
less noisy, with a relative modelling error of about 55\%
(Fig.~\ref{errorpic}). One feature that can never be recovered in
any of the reconstructions is the southern part of the rift, which
does not lie between any station-event pair.

\begin{figure*}
\centering
\resizebox{\textwidth}{!}{\includegraphics{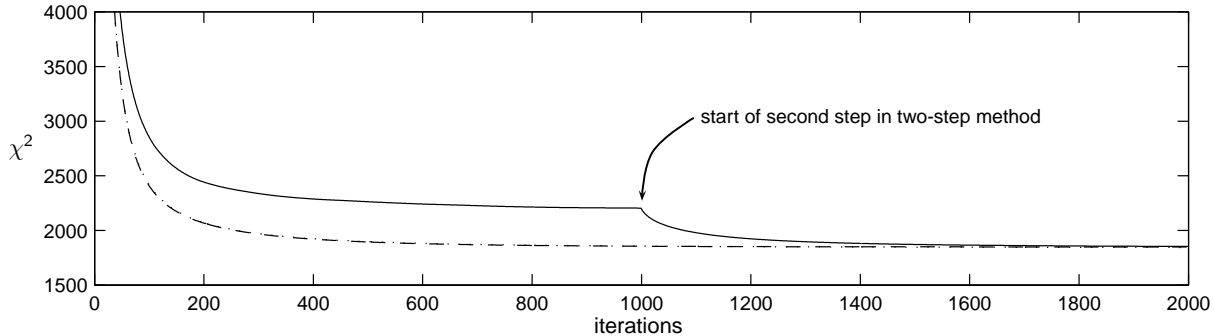}}

\caption{Graph of the variance-adjusted data fit
$\chi^2=\|\data-\A \m\|_2^2\hspace{0.2em}/\sigma^2$
versus the number of
iterations: two-step $\ell_1$-regularization method (solid line), spatial-domain
$\ell_2$ method (dashed line) and wavelet-basis $\ell_2$ method
(dotted line ). The thresholding and penalization parameter
$\tau$ has in each case been tailored so that the final value of $\chi^2$, after
$1000+1000$ or 2000 iterations, is equal to the number of data, namely 1848.
Note the improvement in the rate of convergence toward the model with
$\chi^2=1848$ after the implementation of the second step in the two-step
$\ell_1$ iteration.}
\label{chi2pic}
\end{figure*}

\begin{figure*}
\centering\resizebox{\textwidth}{!}{\includegraphics{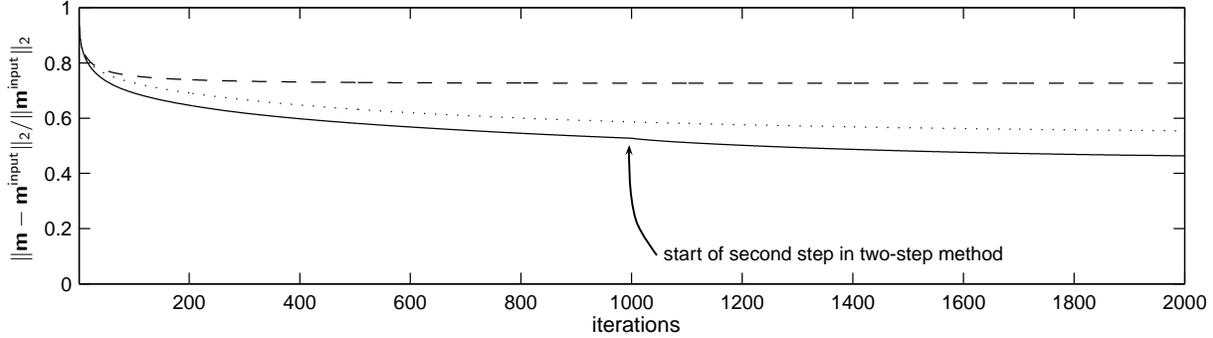}}

\caption{Graph showing the relative modeling error
$\|\m-\m^{\mathrm{input}}\|_2/\|\m^{\mathrm{input}}\|_2$ versus the
number of iterations: two-step $\ell_1$ method (solid line),
spatial-domain $\ell_2$ method (dashed line) and wavelet-basis
$\ell_2$ method (dotted line). The $\ell_1$-regularization method
clearly yields the most faithful reconstruction of the input model
$\m^{\mathrm{input}}$. Note the (slight) improvement in the rate of
decrease of the modelling error following the start of the second
step in the two-step $\ell_1$ iteration.} \label{errorpic}
\end{figure*}

In Fig.\ \ref{waveletpic} we compare the wavelet coefficients $\w$
of the input model, the two-step $\ell_1$ reconstruction and the
wavelet-basis $\ell_2$ reconstruction. In accordance with our basic
assumption, the $\ell_1$-regularized model is sparse in the wavelet
basis. Most of the small-scale coefficients $\w$ are zero ---in
agreement with the original model on the left--- indicating the
effectiveness of the iterative thresholding algorithm. The wavelet
coefficients of the wavelet $\ell_2$ reconstruction are clearly not
sparse. Also this solution seems to suffer from large-scale
artifacts (see Fig.\ \ref{modelfigure}, rightmost map).

\begin{figure*}
\centering\resizebox{\textwidth}{!}{\includegraphics{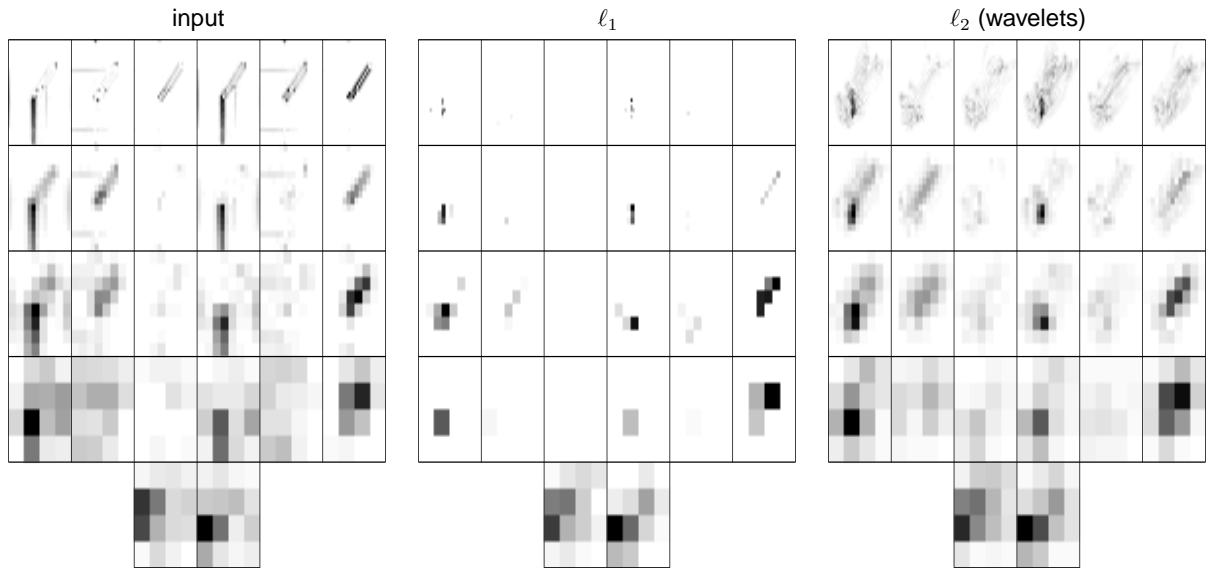}}

\caption{Graphical display of (the modulus of) the wavelet and
scaling coefficients. Left: coefficients of the synthetic input
model $\m^{\mathrm{input}}$. Middle: coefficients of the two-step
$\ell_1$ model after $1000+1000$ iterations. Right: coefficients of
the wavelet-basis $\ell_2$ model after 2000 iterations. The four
wavelet scales are plotted, smallest to largest, top to bottom. Each
row shows the six different wavelet directions, plotted next to each
other in the same left-to-right order as the wavelets plotted in
Fig.~\ref{waveletpicb}. Scaling coefficients are plotted on the
bottom row. White denotes a zero coefficient, $w_i=0$. Each
rectangle corresponds to the spatial domain $25\degr$E -- $50\degr$E
by $15\degr$S -- $20\degr$N.} \label{waveletpic}
\end{figure*}

\begin{figure*}
\centering
\resizebox{\textwidth}{!}{\includegraphics{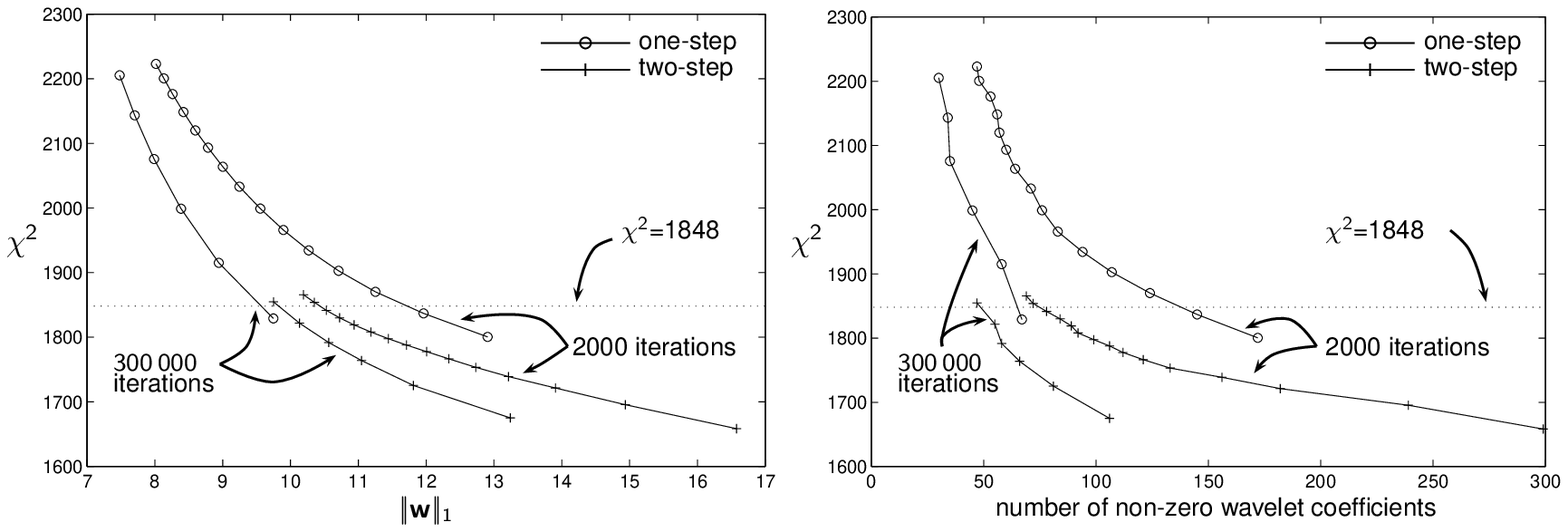}}

\caption{Left: Data misfit $\chi^2$ versus $\ell_1$ wavelet norm
$||\w||_1$ tradeoff curve. Right: Alternative tradeoff curve showing
$\chi^2$ versus the number of nonzero wavelet coefficients of the
model $\m$. Different values of the thresholding parameter $\tau$
were used to determine each point on the various curves. Circles:
one-step method, after either 2000 or $300\,000$ iterations;
crosses: two-step method after an equivalent number (either
$1000+1000$ or $150\,000+150\,000$) of total iterations. The
relative positions of the one-step and two-step curves suggests that
$300\,000$ total iterations is sufficient to achieve full
convergence. The horizontal dotted lines show the statistically
meaningful value of the noisy data misfit, $\chi^2=1848$ (the number
of data).} \label{tradeoffpic}
\end{figure*}

In the leftmost plot in Fig.~\ref{tradeoffpic} we show $\chi^2$
versus $\|\w\|_1$ tradeoff curves for the $\ell_1$ reconstruction
method, both with and without using the two-step procedure. After
$1000+1000$ iterations, the $\ell_1$ wavelet norm $\|\w\|_1$ of the
two-step reconstructed model is lower --- for the same value of
$\chi^2$ --- than the corresponding norm of the model produced by
2000 iterations of  the first step, with no subsequent redefinition
of the data $\data$ and reiteration. This is an indication that 2000
total iterations is inadequate to achieve full convergence, since
the fully converged model, which minimizes the functional $I_1(\w)$
given in eq.~(\ref{l1functional}), must be the minimum-norm model
for a fixed value of $\chi^2$ by definition. A much larger number of
iterations seems to be required to guarantee convergence. To
construct the second set of tradeoff curves in
Fig.~\ref{tradeoffpic}, we employed $150\,000+150\,000$ iterations
in the two-step case and $300\,000$ in the single-step case; such a
large number would be prohibitive in any larger-scale, more
realistic, 3D application. We have chosen to limit the iteration
counts to $1000+1000$ or 2000 in all of our model-space comparisons,
since any changes in the spatial-domain features of the models $\m$
are barely discernible to the eye with further iteration. The
rightmost plot in Fig. \ref{tradeoffpic} shows the principal
advantage of using the two-step iteration procedure: for the same
total number of iterations, either $1000+1000=2000$ or
$150\,000+150\,000=300\,000$, the number of nonzero wavelet
coefficients of the two-step models is always lower than the
corresponding number for the single-step models. The two-step
$\ell_1$ procedure therefore leads more quickly to a sparser
wavelet-basis solution, as expected.

We also compared the single-step and two-step $\ell_1$ inversion methods
with the corresponding $\ell_2$ reconstruction methods, both with and without
wavelets, for a number of other input synthetic models.
These include three checkerboard patterns of decreasing
scale and a model similar to the geologically inspired one
in Fig. \ref{modelfigure}, but with a more curvaceous low-velocity rift
(see Fig. \ref{allpics}). Both the single-step $\ell_1$ reconstructions and
the $\ell_2$ reconstructions are computed using 2000 iterations, whereas the
two-step $\ell_1$ models are computed using $1000+1000$ iterations.
In all cases, the two-step $\ell_1$ models are the most
parsimonious and therefore to most geoscientists the most
acceptable. One could consider using smoothness damping
to improve the quality of the $\ell_2$ images; however, this
would be done at the cost of resolving the sharpness of the rift
structure.  A nitpicker could perhaps also argue that the ``rift'' structure in the model
produced by the $\ell_1$ procedure extends further northwards,
albeit diminished in amplitude,
whereas conventional $\ell_2$ regularization without wavelets exhibits a
sharper cutoff, more like the input model. It achieves this sharp cutoff,
however, at the expense of many artifacts elsewhere, especially
along dominant ray directions. Perhaps the most noteworthy
feature of the $\ell_1$ regularization method is its suppression of
the artifacts resembling high-frequency kernel images that are
streaked along surface-wave raypaths in all the $\ell_2$ models,
to the north of the rift and within the craton. This is one of
the most serious artifacts that plague conventional seismic
tomography: $\ell_2$ regularization frequently if not always
seems to enhance the well-sampled regions of the model.
The $\ell_1$ wavelet-basis reconstructions show no signs of this
familiar deficiency.

The computational bottleneck in the present 2D synthetic study is
not the wavelet transform --- which is fast, certainly on a model
$\m$ of modest dimension $64\times 64$ --- or even the number of
iterations, but it is simply the size of the matrix $\A $. A
significant amount of time is needed to accurately pre-compute $\A
$, and considerable memory is needed to store the computed elements
in memory; this is necessary because the product $\A ^\tr \A $ is
used in every step of the iteration. Doubling of the resolution in
every direction results in a fourfold increase in size of the model
$\m$, and a sixteen-fold increase in the number of elements in the
square matrix $\A ^\tr \A$. All calculations were performed using
\textsc{Matlab}; software for the 2D dual-tree wavelets was
downloaded from \cite{selesnick06}.

\begin{figure*}
\center\resizebox{\textwidth}{!}{\includegraphics{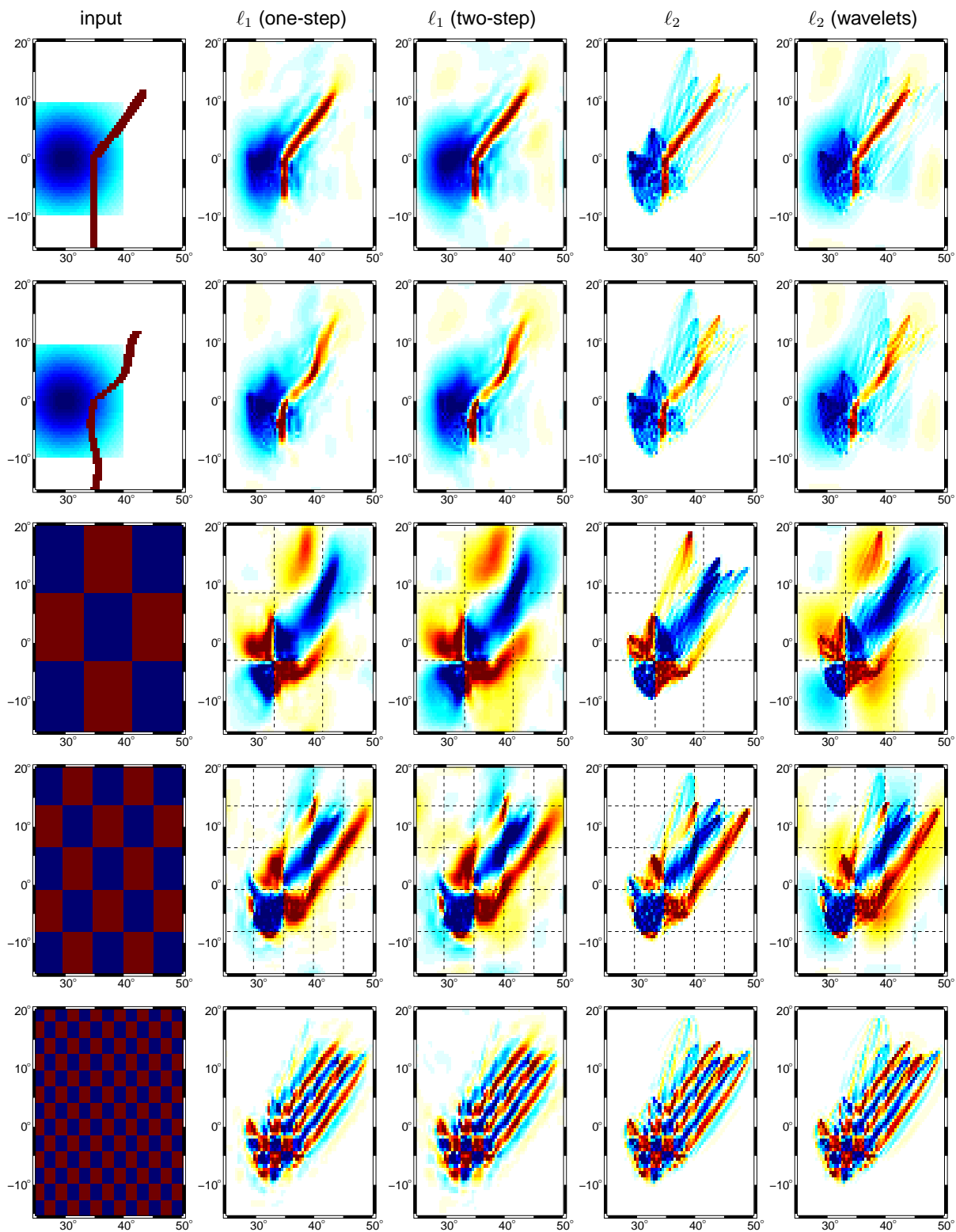}}

\caption{Results of applying different reconstruction techniques to a number of different 2D
toy models. From left to right: Original input model; $\ell_1$ reconstruction
(2000 iterations, single-step procedure); $\ell_1$ reconstruction
($1000+1000$ iterations, two--step procedure); $\ell_2$ reconstruction
(2000 iterations, without wavelets); $\ell_2$ reconstruction (2000 iterations,
with wavelets).} \label{allpics}
\end{figure*}

\section{Conclusions}

We tested several new methods of regularization through wavelet
decomposition of a toy 2D tomographic problem characterized by both
smooth and sharp velocity anomalies. A variety of synthetic
inversion experiments show that minimization of the $\ell_1$-norm of
a wavelet decomposition of the model leads to tomographic images
that are parsimonious in the sense that they use only a few wavelets
and still  represent both smooth and sharp features well without
introducing significant blurring or artifacts. The $\ell_1$-norm
performs significantly better than an $\ell_2$ regularization on
either the model or its wavelet decomposition. In particular,
raypath-associated artifacts are almost completely suppressed.

The choice of dual-tree complex wavelets in 2D, representing six
space directions, is sufficient to avoid directional bias, and
efficient in modeling both smooth features such as the cratonic
structure as well as sharp features such as the rift structure in
our simplified synthetic model. Numerical comparisons between the
inversion results and the input model used to generate the data
confirm the superiority of the $\ell_1$-norm regularization. Though
in real-world inversions such ground-truth information is not
available, one can argue that the $\ell_1$ inversion method serves
the principle of parsimony well and is to be preferred over more
common methods.  If the tomographic object (such as the real earth)
is too complex to be well represented by a parsimonious expansion in
wavelets, neither method is able to resolve such complexity
adequately with a limited data set, as shown in the bottom rows of
Fig. \ref{allpics}, where even the $\ell_1$ inversions begin to show
the effects of raypath distribution. In this case, we expect that
the principle of parsimony can be usefully applied once a richer
family of building blocks is considered.

The only drawback of the method, so far, is the slow convergence of
the $\ell_1$ surrogate-functional iteration procedure. Our
preference for the thresholded algorithm used here arises from the
fact that its convergence is guaranteed even though the $\ell_1$
problem is nonlinear. We have introduced a two-step procedure that
leads to a significant speedup; however, Fig.~\ref{tradeoffpic}
indicates that even $1000+1000$ iterations do not suffice for
complete convergence (it nevertheless produces an excellent
approximation). A potentially promising approach towards further
convergence improvement is to combine an efficient linear method
(such as e.g. conjugate-gradient) with an \emph{adaptive}
thresholding scheme. This would then avoid the need to precompute
the largest eigenvalue of $\A ^\tr \A$ and facilitate the
application of the $\ell_1$ method to a larger, 3D, study of
body-wave tomography.

\section{Acknowledgments}

Financial support for this work was provided by NSF grant DMS-0530865.
I.L. is a postdoctoral fellow with the F.W.O.-Vlaanderen (Belgium).

\appendix

\section{Two-dimensional sensitivity kernels}
\label{kernelappendix}

\renewcommand{\theequation}{\thesection\arabic{equation}}
\setcounter{equation}{0}

The toy linear inverse problem $\A\m = \data$ used in this paper is
designed to incorporate all the important characteristics of a
real-world regional tomographic inversion, while at the same time
being small enough to allow for repeated experimenting with
reasonable CPU times on a single workstation. For this reason, we
limit attention to surface-wave dispersion data, specifically
perturbations $\delta k (\nu)$ in the wavenumber $k(\nu)$, presumed
to be measured in rad/m, of the fundamental ($n=0$) Rayleigh mode at
temporal frequency $\nu$, measured in Hz. Finite-frequency theory
based upon the Born approximation \cite{zhou04} gives a linear
relationship between such wavenumber perturbations and the 3D
perturbations in the fractional shear-wave velocity
$\delta\hspace{-0.1em}\ln\hspace{-0.1em}\beta(\bx)$ within the
earth:
\begin{equation}
\delta k (\nu ) = \int\!\!\!\int\!\!\!\int K_{\rm 3D}(\bx,\nu )
\,\delta\hspace{-0.1em}\ln\hspace{-0.1em}\beta(\bx )\,\, \dd^3 \bx
\label{3d}.
\end{equation}
Making use of a number of flat-earth approximations that do not
fundamentally affect the nature of the inverse problem, we can write
the 3D Fr\'{e}chet sensitivity kernel $K_{\rm 3D}(\bx,\nu)$, for the
simplest case of an explosive source with an isotropic radiation
pattern and a measurement made on the vertical component at the
receiver, in the form \iftwocol{\begin{equation} \begin{array}{l}
\displaystyle K_{\rm 3D} (\bx ,\nu )\! =\! \left [ e_{0} (z,\nu )+\!
e_{1} (z,\nu ) \cos
\eta\! + e_{2} (z,\nu ) \cos 2\eta \right ]\\
\displaystyle\quad\qquad\qquad\qquad \times\left ( \frac{ 1}{8 \pi
k(\nu )l\,l^\prime
l^{\prime\prime}}\right )^\frac12\\
\displaystyle\quad\qquad\qquad\qquad\qquad
\times\sin\!\left[k(\nu)(l' + l^{\prime\prime} - l +\pi /4)\right],
\end{array}\label{Jwx}
\end{equation}}{\begin{equation} K_{\rm 3D} (\bx
,\nu ) = \left [ e_{0} (z,\nu )+ e_{1} (z,\nu ) \cos \eta + e_{2}
(z,\nu ) \cos 2\eta \right ] \left ( \frac{ 1}{8 \pi k(\nu
)l\,l^\prime l^{\prime\prime}}\right )^\frac12 \sin\!\left[k(\nu)(l'
+ l^{\prime\prime} - l +\pi /4)\right], \label{Jwx}
\end{equation}}
where $z$ is the depth, $l$ is the epicentral distance measured in m
on the surface of the earth, and $l'$ and $l^{\prime\prime}$ are the
horizontal distances of the scatterer $\bx=(x,y,z)$ from the source
and receiver, respectively. The quantity $\eta$ is the scattering
angle, measured at the surface projection $(x,y)$ of $\bx$, as shown
in Figure~\ref{f1}. Expressions for the depth-dependent functions
$e_{0}(z,\nu )$, $e_1(z,\nu)$ and $e_2(z,\nu)$ can be found in the
appendix of \cite{zhou04}.

\begin{figure}
\centering\includegraphics{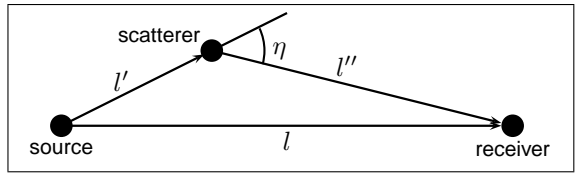}

\caption{Schematic map view of the single-scattering geometry in our
simplified 2D, flat-earth, surface-wave inversion problem. The
quantity $l$ is the horizontal epicentral distance between the
source and receiver; $l'$ and $l^{\prime\prime}$ are the lengths of
the first and second legs of the detour path, respectively, and the
angle $\eta$ measures the deflection of the wave at the scatterer.}
\label{f1}
\end{figure}

To simplify matters even further, we assume that the velocity
perturbation $\delta\hspace{-0.1em}\ln\hspace{-0.1em}\beta(\bx)$ is
independent of depth $z$ and dependent only upon the horizontal
Cartesian coordinates $x$ and $y$. Upon integrating the factors
$e_{0}(z,\nu )$, $e_1(z,\nu)$ and $e_2(z,\nu)$ over depth,
\iftwocol{\begin{equation}\begin{array}{l} \displaystyle E_{0} (\nu
) = \int_0^\infty e_{0} (z, \nu ) \,\dd z,\qquad E_{1} (\nu ) =
\int_0^\infty e_{1} (z, \nu ) \, \dd z,\\[4mm]  \displaystyle E_{2} (\nu
) = \int_0^\infty e_{2} (z, \nu ) \, \dd z,\end{array}
\end{equation}}{\begin{equation} E_{0} (\nu ) = \int_0^\infty e_{0} (z,
\nu ) \,\dd z,\qquad E_{1} (\nu ) = \int_0^\infty e_{1} (z, \nu ) \,
\dd z,\qquad E_{2} (\nu ) = \int_0^\infty e_{2} (z, \nu ) \, \dd z,
\end{equation}} we may then relate $\delta k(\nu)$ to
$\delta\hspace{-0.1em}\ln\hspace{-0.1em}\beta(x,y)$ via a 2D
sensitivity kernel:
\begin{equation}
\delta k (\nu ) = \int\!\!\!\int K_{\rm 2D}(x,y,\nu )\,
\delta\hspace{-0.1em}\ln\hspace{-0.1em}\beta(x,y ) \, \dd x\, \dd y \label{2d},
\end{equation}
where\iftwocol{\begin{equation}\begin{array}{l} K_{\rm 2D} (x,y ,\nu
) = \left [ E_{0} (\nu)+ E_{1} (\nu ) \cos \eta + E_{2} (\nu ) \cos
2\eta \right ]\\ \quad\qquad\qquad\qquad\displaystyle\times\left (
\frac{ 1}{8 \pi k(\nu ) l\,l^\prime
l^{\prime\prime}}\right )^\frac12 \\
\displaystyle\quad\qquad\qquad\qquad\qquad\times\sin\!\left[k(\nu)(l'
+ l^{\prime\prime} - l +\pi /4)\right]. \label{Jwx2}
\end{array}\end{equation}}{\begin{equation} K_{\rm 2D} (x,y ,\nu ) = \left [
E_{0} (\nu)+ E_{1} (\nu ) \cos \eta + E_{2} (\nu ) \cos 2\eta \right
] \left ( \frac{ 1}{8 \pi k(\nu ) l\,l^\prime
l^{\prime\prime}}\right )^\frac12 \sin\!\left[k(\nu)(l' +
l^{\prime\prime} - l +\pi /4)\right]. \label{Jwx2}
\end{equation}}
The rapidly oscillating sinusoidal function $\sin [k(\nu)(l' +
l^{\prime\prime} - l +\pi /4)]$ in eq.~(\ref{Jwx2}) is constant on
ellipses, $l' + l^{\prime\prime}=\mbox{constant}$, having the
surface projections of the source and receiver as foci. The
$\cos\eta$ and $\cos 2\eta$ dependence and the term involving the
integrable singularity $1/\sqrt{l' l^{\prime\prime}}$ act to slowly
modulate this dominant elliptical dependence.

Eqs~(\ref{2d}) and~(\ref{Jwx2}) are valid, subject to the already
noted approximations, for a monochromatic wavenumber perturbation
$\delta k(\nu)$, whereas actual surface-wave dispersion measurements
must of necessity be made on a portion of a seismogram of finite
length, typically multiplied by a time-domain taper $h(t)$.
\cite{zhou04} show that the effect of such a finite-length taper can
be accounted for by  modifying the taper as follows:
\begin{equation} \label{tony2}
K_{\rm 2D}(x,y,\nu ) \rightarrow K_{\rm 2D}(x,y,\nu )
\,h((l' +l^{\prime\prime})/C(\nu )),
\end{equation}
where $C(\nu )$ is the the group velocity at frequency $\nu$
measured in m/s. This modification has the effect of limiting the
cross-path width of the Fr\'echet kernel $K(x,y,\nu)$, since
$h(t)=0$ for large detour times.  We assume the data $\delta k(\nu)$
have been measured using a Hann or cosine taper, of duration five
wave periods centered on the group arrival time:
\iftwocol{\begin{equation} \label{tony3}
h(t)=\left\{\begin{array}{l}
0  \\
\frac{1}{2}[1-\cos2\pi\nu(t-t_{\mathrm{arrival}}-2.5/\nu)]
\\
0
\end{array}\right.
\end{equation}for $t\leq t_{\mathrm{arrival}}-2.5/\nu$, for
$t_{\mathrm{arrival}}-2.5/\nu\leq t\leq
t_{\mathrm{arrival}}+2.5/\nu$ and for $t\geq
t_{\mathrm{arrival}}+2.5/\nu$ respectively, and}{\begin{equation}
\label{tony3} h(t)=\left\{\begin{array}{ll}
0 &\quad\mathrm{for}\quad t\leq t_{\mathrm{arrival}}-2.5/\nu\\
\frac{1}{2}[1-\cos2\pi\nu(t-t_{\mathrm{arrival}}-2.5/\nu)] &
\quad\mathrm{for}\quad
t_{\mathrm{arrival}}-2.5/\nu\leq t\leq t_{\mathrm{arrival}}+2.5/\nu\\
0 & \quad\mathrm{for}\quad t\geq t_{\mathrm{arrival}}+2.5/\nu
\end{array}\right.
\end{equation}}
where $t_{\mathrm{arrival}}=l/C(\nu)$. Since $l' +
l^{\prime\prime}\geq l$ only the $t\geq t_{\mathrm{arrival}}$
portion of the taper~(\ref{tony3}) contributes to the
finite-record-length sensitivity kernel~(\ref{tony2}).

The group velocity $C(\nu)$, unperturbed wavenumber $k(\nu)$ and
auxiliary variables $E_0(\nu)$, $E_1(\nu)$ and $E_2(\nu)$ for
fundamental-mode Rayleigh waves are listed in Table~\ref{datatable}
at the eight selected frequencies $\nu$; the corresponding wave
periods vary roughly between 100 and 10 s. Since $E_0(\nu)$,
$E_1(\nu)$ and $E_2(\nu)$ are all negative, a positive velocity
perturbation, $\delta\hspace{-0.1em}\ln\hspace{-0.1em}\beta(x,y)>0$,
gives rise to a negative wavenumber perturbation, $\delta k(\nu)<0$,
i.e.\ an apparently longer wavelength wave, as expected. See
Fig.~\ref{coveragepic} for two examples of sensitivity kernels
$K_{\rm 2D}(x,y,\nu)$ computed in this way. It is noteworthy that a
2D surface-wave inversion based upon eqs~(\ref{2d})--(\ref{tony3})
differs from the common approach of inverting for a 2D phase
velocity map at a single specified frequency $\nu$: such maps are
strictly incompatible with the notion of finite frequency, where no
local phase velocity can be defined except when very crude
approximations are made; for a discussion of this issue see
\cite{zhou04}.

\begin{table*}
\caption{Parameters $\nu, C(\nu), k(\nu), E_0(\nu), E_1(\nu)$ and
$E_2(\nu)$ needed to compute the simplified 2D sensitivity kernels
$K_{\rm 2D}(x,y,\nu)$. Fundamental-mode Rayleigh-wave measurements
$\delta k(\nu)$ are presumed to have been made at eight frequencies
ranging between $\nu\approx 0.01$~Hz (100~s period) and $\nu\approx
0.1$~Hz (10~s period).} \centering
\begin{tabular}{|r@{.}lr@{.}lr@{.}lr@{.}lr@{.}lr@{.}l|}
\hline   \multicolumn{2}{c}{\rule[-1.5mm]{0pt}{5.5mm}$\nu$\,(mHz)} &
\multicolumn{2}{c}{$C$\,(m/s)} &
\multicolumn{2}{c}{$k\,(10^{-4}\mathrm{m}^{-1})$} &
\multicolumn{2}{c}{$E_{0}\,(10^{-9}\mathrm{m}^{-2})$} &
   \multicolumn{2}{c}{$E_{1}\,(10^{-9}\mathrm{m}^{-2})$}
   & \multicolumn{2}{c}{$E_{2}\,(10^{-9}\mathrm{m}^{-2})$}\\
\hline
  $10$&$742$ & $3831$&$3$ & $0$&$165\,37$ & $-0$&$079\,642$ & $-0$&$359\,72$ & $-0$&$061\,743$\\
  $15$&$625$ & $3829$&$8$ & $0$&$245\,11$ & $-0$&$126\,90$  & $-0$&$776\,64$ & $-0$&$129\,84$\\
  $20$&$508$ & $3751$&$3$ & $0$&$325\,77$ & $-0$&$176\,86$  & $-1$&$365\,6$  & $-0$&$222\,47$\\
  $30$&$273$ & $3434$&$4$ & $0$&$495\,75$ & $-0$&$368\,58$  & $-3$&$217\,7$  & $-0$&$489\,28$\\
  $40$&$039$ & $3064$&$8$ & $0$&$684\,98$ & $-1$&$078\,3$   & $-6$&$216\,1$  & $-0$&$944\,65$\\
  $50$&$781$ & $2861$&$6$ & $0$&$914\,30$ & $-2$&$788\,5$   & $-10$&$769$  & $-1$&$815\,0$\\
  $70$&$313$ & $2872$&$8$ & $1$&$344\,6$  & $-6$&$417\,5$   & $-22$&$322$  & $-4$&$130\,0$\\
  $99$&$609$ & $2971$&$5$ & $1$&$973\,3$  & $-11$&$684$     & $-47$&$879$  & $-8$&$884\,8$\\
\hline
\end{tabular}
\label{datatable}
\end{table*}

\section{Notes on wavelets}

\renewcommand{\thefigure}{\thesection\arabic{figure}}
\setcounter{figure}{0}

\label{waveletappendix}


The basic building block of the 1D discrete wavelet transform (DWT)
is a filter bank. It consists of a high-pass filter $g$ (i.e. a
generalized difference) and a low-pass filter $h$ (i.e. a
generalized average) that are applied to a given signal $\m$ (i.e. a
list of numbers) in the following way: $\m$ is convolved with $g$
and downsampled, $\m$ is convolved with $h$ and downsampled. This
results in two signals, each with half the length of the original
one. The process can be inverted by upsampling (inserting zeroes)
the two resulting sequences and convolving each with two (carefully
matched) filters $\tilde g$ and $\tilde h$ and then adding the two.
A traditional way of representing this procedure is shown in the
left of Fig.~\ref{filterbankpic}. It turns out that there exist
\emph{finite} filters that give rise to perfect reconstruction
(these use finite convolutions only and lead to compactly supported
wavelets); moreover in some very special cases, one can have that
finite $\tilde g$ and $\tilde h$ are the reverse of $g$ and $h$
(corresponding to compactly supported \emph{orthogonal} wavelets).
The Haar wavelets have $g=(\frac{1}{2},-\frac{1}{2})$ and
$h=(\frac{1}{2},\frac{1}{2})$, but there exist longer (perfect
reconstruction) finite filters (which give rise to smoother
wavelets). The so-called D4 wavelets correspond to $h=(1+\sqrt{3},
3+\sqrt{3}, 3-\sqrt{3}, 1-\sqrt{3})/4\sqrt{2}$ and $g=(1-\sqrt{3},
-3+\sqrt{3}, 3+\sqrt{3},-1-\sqrt{3})/4\sqrt{2}$.

\begin{figure*}
\centering\resizebox{\textwidth}{!}{\includegraphics{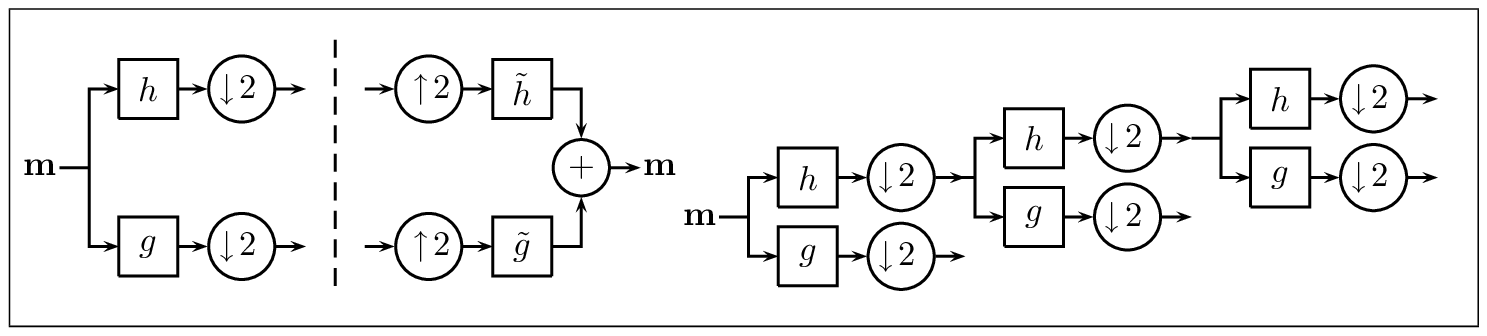}}

\caption{Left: Schematic representation of a perfect-reconstruction
filter bank that can be used to decompose or reconstruct a 1D signal $x$.
Right: A standard wavelet tree.}\label{filterbankpic}
\end{figure*}

The 1D discrete wavelet transform is defined by the iteration of the
analysis filter bank on the low-pass outcomes (see right side of
Fig.~\ref{filterbankpic}). In this way, successive levels of detail
are stripped of the input signal $\m$ (and stored in wavelet
coefficients), leaving a very coarse average (stored in so-called
scaling coefficients). This construction is called a wavelet tree.
It not only defines the DWT but also provides its practical
implementation. When using finite filters, the construction
automatically gives rise to a computationally efficient algorithm:
as a result of the subsampling each step cost only half as much time
as the previous one. The total number of operations then is
$kN+kN/2+kN/4+kN/8+\ldots=2kN$, less than the $\mathcal{O}(N^2)$ for
a generic linear transformation.

A standard way of generating wavelets in 2D is to form the direct
product of 1D wavelets, i.e. the filters are applied to rows and
columns of an image (lo-lo, lo-hi, hi-lo and hi-hi). This, however,
has the marked disadvantage of poor directional sensitivity. In this
study, to obtain better directional sensitivity, we use the complex
2D wavelets developed by \cite{Kingsbury}. These are constructed
also by direct product but from \emph{two} simultaneous wavelet
trees (see \cite{Kingsbury2} for a diagram of such a dual tree). The
qualitative difference between these two constructions is best seen
in the Fourier domain. Fig.~\ref{fourierpic} shows a schematic
representation of the supports of the Fourier transforms of the
wavelet functions, both for the usual 2D wavelet construction and
for the 2D complex wavelets. The two are fundamentally different:
whereas the usual separable 2D wavelet construction gives rise to a
horizontal, a vertical and one (!) diagonal part at each scale, the
complex 2D construction has six different inherent directions per
scale. A careful choice of the different filters also leads to an
(almost) tight frame (i.e. the inverse wavelet transform almost
coincides with the transpose).

\begin{figure*}
\centering
\resizebox{\textwidth}{!}{\includegraphics{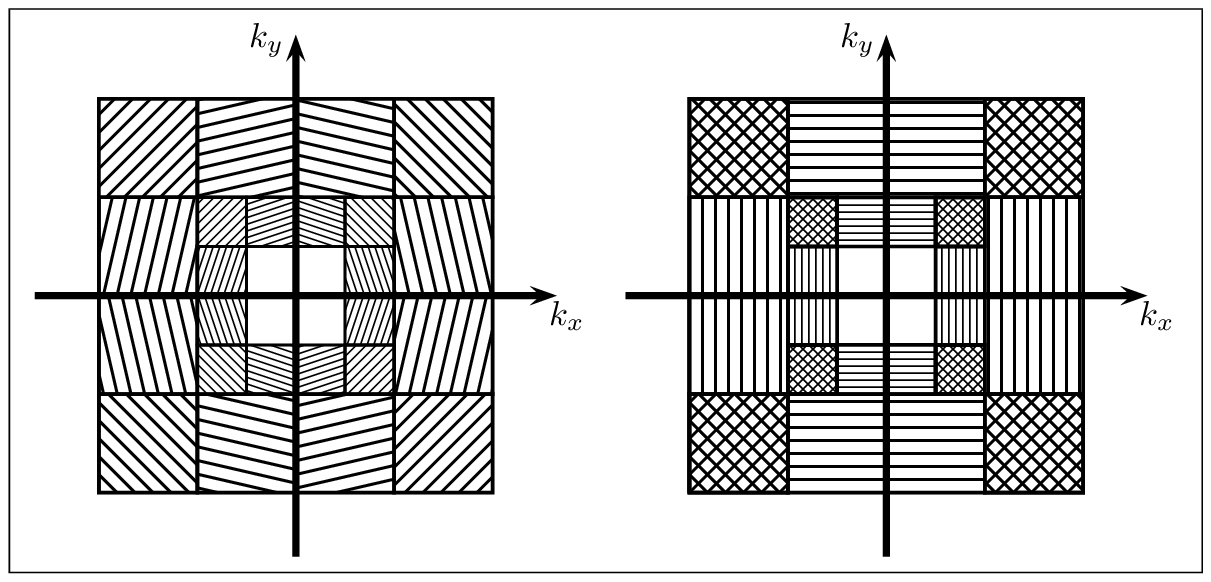}}

\caption{Partitioning of the 2D Fourier domain $(k_x,k_y)$ by the
supports of the Fourier transform of wavelet functions. Only two
wavelet scales are shown with the finest one on the outside. In
practice the supports have smooth (overlapping) tapers. Left: With
the complex 2D wavelets used in this paper, all squares come in
pairs giving rise to six dominant directions (as indicated by the
hatch patterns). Right: The standard (direct product) 2D
construction only has horizontal and vertical sensitivity; the
`corner' (hi-hi) squares encode both $45\degr$ and $-45\degr$ at the
same time.}\label{fourierpic}
\end{figure*}

The price to pay for these benefits is the redundancy. In 2D the
complex wavelets generate four times as many coefficients as there
are pixels in the original image (two trees and real and imaginary
parts of the output). E.g. the $64\times 64$ spatial-domain images
we use in Section~\ref{implementationsection} give rise to $16\,320
=2\times 6\times (32^2+16^2+8^2+4^2)$ wavelet coefficients and $64
=2\times 2\times 4^2$ scaling coefficients (see e.g.
Fig.~\ref{waveletpic}). Together this is $16\,384$ which equals
$4\times 64^2$.

\end{document}